\documentclass[12pt,preprint]{aastex}
\usepackage{pdflscape}
\shorttitle{LONG-PERIOD VARIATIONS IN THE RADIAL VELOCITY OF SPECTROSCOPIC BINARY M GIANT $\mu$ URSAE MAJORIS}
\shortauthors{Lee et al.}

\begin{document}

\title{LONG-PERIOD VARIATIONS IN THE RADIAL VELOCITY OF SPECTROSCOPIC BINARY M GIANT $\mu$ URSAE MAJORIS}
\author{Byeong-Cheol Lee\altaffilmark{1,2}, Inwoo Han\altaffilmark{1}, Myeong-Gu Park\altaffilmark{3}, David E. Mkrtichian\altaffilmark{4,5}, Artie P. Hatzes\altaffilmark{6}, Gwanghui Jeong\altaffilmark{1,2}, and Kang-Min Kim\altaffilmark{1}}

\altaffiltext{1}{Korea Astronomy and Space Science Institute 776, Daedeokdae-ro, Yuseong-gu, Daejeon 305-348, Korea; bclee@kasi.re.kr}
\altaffiltext{2}{Astronomy and Space Science Major, Korea University of Science and Technology, Gajeong-ro Yuseong-gu, Daejeon 305-333, Korea}
\altaffiltext{3}{Department of Astronomy and Atmospheric Sciences, Kyungpook National University, Daegu 702-701, Korea}
\altaffiltext{4}{National Astronomical Research Institute of Thailand, Chiang Mai 50200, Thailand}
\altaffiltext{5}{Crimean Astrophysical Observatory, Taras Shevchenko National University of Kyiv, Nauchny, Crimea, 98409, Ukraine}
\altaffiltext{6}{Th{\"u}ringer Landessternwarte Tautenburg (TLS), Sternwarte 5, 07778 Tautenburg, Germany}

%---------------------------------------------------------------------------------
\begin{abstract}
We report that the spectroscopic binary $\mu$ Ursae Majoris ($\mu$ UMa) has secondary RV variations of 471.2 days in addition to those of 230.0 days already known. Keplerian orbit analysis yields stellar mass companions of 1.6~$M_{\odot}$ for the 230-d period and 0.14~$M_{\odot}$ for the 471-d period. However, the \emph{HIPPARCOS} photometries show a period similar to the stellar rotational period, which is one-quarter of the RV period. Variations in the bisector velocity curvature show a period of 463.6 days. We also find $\sim$473-day variations in the equivalent width (EW) measurements of the H$_{\alpha}$ and H$_{\beta}$ lines, whose origin is probably stellar activity. We note that the nature of 471-day variations is similar to one observed in ``Sequence D'' of Asymptotic Giant Branch (AGB) pulsating stars. We therefore conclude that the RV and the EW variations in the spectroscopic binary M giant $\mu$ UMa~A originate from the complex pulsations and the chromospheric activity.
\end{abstract}


\section{INTRODUCTION}
Low-amplitude  radial velocity (RV) variations are common among  evolved K and M giant stars
and these have both long (hundreds of days) and short (hours to days) time scales. Short-term variations are most likely caused by Sun-like oscillations  excited by turbulent convection and by convective motions in the outer convective zone. Long-term variations can arise from either sub-stellar companions, surface inhomogeneities, pulsations, or other intrinsic stellar mechanisms.
For giant stars, the size of surface granulation is linearly dependent on the stellar radius and mass \citep{fre02}. Thus, the atmosphere of red giants should exhibit small number of large cells that may cause stochastic low-amplitude RV or light variations. These different time-scale variations can occur in K and M giant stars simultaneously and may have complex influence in the observed RVs of stars \citep{blu94,hat98,lar99}. Long-term precise RV monitoring of giant stars is an important tool to study the origin of these variations.

In 2006, we initiated a precise RV survey for 10 bright  M giants as a part of ongoing K giant exoplanet survey using the 1.8 m telescope at Bohyunsan Optical Astronomy Observatory (BOAO). The M giant $\mu$ UMa is one star in our sample for which we have obtained precise RV measurements over the past eight years. Here we report the detection of long-period RV variations for $\mu$~UMa, possibly caused by rotational modulation of surface inhomogeneities. In Sect. 2, the stellar characteristics of the host star are derived. In Sect. 3, we describe our observations and data analysis. The RV variation measurements and possible origins are presented in Sect. 4. Finally, in Sect. 5, we discuss and summarize our findings.

%________________________________________________________________

\section{THE STAR $\mu$ URSAE MAJORIS}
\subsection{Fundamental parameters}
$\mu$ UMa is a well-studied, single-lined spectroscopic binary system (SB1) with a companion at a distance $\sim$1 AU from the primary with an orbital period of $\sim$230 days \citep{jac57,luc71,eat07}. It is an evolved star that is currently in the red giant stage with a stellar classification of M0~III. An apparent visual V-band magnitude of 3.15 places it among the brighter members of the constellation Ursa Major. The \emph{HIPPARCOS} parallax \citep{van07} gives a distance of 76 pc, yielding an absolute V-band magnitude of $-$1.09. It has expanded to 74.7 times the radius of the Sun while the outer atmosphere has cooled to an effective temperature of 3899 K. The estimated luminosity is 1148 $L_{\odot}$, but the star varies in brightness from 2.99 to 3.33 mag. \citet{mas08} estimated a stellar mass of 2.2 $M_{\odot}$ for $\mu$ UMa.  They also measured a rotational velocity of 7.5 ~km~s$^{-1}$ using the cross-correlation technique on the observed spectra against templates drawn from a library of synthetic spectra calculated by \citet{kur92} for different stellar atmospheres.

\subsection{Photocentric orbital determination}
In order to determine the companion mass in spectroscopic binaries, a complementary technique sensitive to the inclination is required.
In general, most orbital parameters can be derived from RV measurements: a period (\emph{P}), an eccentricity (\emph{e}), an RV semi-amplitude (\emph{K}), a periastron time (\emph{T$_{\rm{periastron}}$}), a longitude of periastron (\emph{$\omega$}), and a primary orbital semi-major axis (\emph{a}). The remaining two orbital parameters, inclination (\emph{i}) and the longitude of the ascending node ($\Omega$), are related to the orientation of the orbit in space.

By combining stellar RV curves and the \emph{HIPPARCOS} astrometric measurements, \citet{ren13} determined the orbits and inclinations of 72 SB1s from $\sim$1200 such binaries in the \emph{HIPPARCOS} catalog, including $\mu$ UMa. The estimated semi-major axis of the primary orbit of $\mu$~UMa is 2.8 $\pm$ 0.2 mas and the orbital inclination is 13.6 $\pm$ 12.8\,$^{\circ}$. Table~\ref{tab1} summarizes the basic stellar parameters for $\mu$~UMa.

%__________________________________________________________________

\section{OBSERVATIONS AND ANALYSIS}
Observations were carried out using the fiber-fed high-resolution Bohyunsan Observatory Echelle Spectrograph (BOES) attached to the 1.8 m telescope at BOAO in Korea. One exposure with the BOES has a wavelength coverage  3500 ${\AA}$ to 10 500 ${\AA}$ distributed over $\sim$80 spectral orders. In order to provide precise RV measurements, we used the 80 $\mu$m diameter fibre which yields a  resolving power $\emph{R}$ = 90 000. The BOES is equipped with an iodine absorption (I$_{2}$) cell needed for more precise RV measurements. Before starlight enters the fiber, it passes through the I$_{2}$ absorption cell regulated at 67\,$^{\circ}$C, which superimposes thousands of molecular absorption lines over the object spectra in the spectral region between 4900 and 6100 ${\AA}$. Using these lines as a wavelength standard, we simultaneously model the time-variant instrumental profile and Doppler shift relative to an I$_{2}$ free template spectrum.

Over the eight-year period from November 2006 to November 2014 (56 nights in total), 112 spectra for $\mu$ UMa were collected. The estimated signal-to-noise (S/N) ratio in the I$_{2}$ region was about 250 with a typical exposure time ranging from 60 to 480 s. The standard reduction procedures of flat-fielding, scattered light subtraction, and order extraction from the raw CCD images were carried out using the IRAF software package. Precise RV measurements using the I$_{2}$ method were undertaken using the RVI2CELL \citep{han07}, which is based on a method by \citet{but96} and \citet{val95}. We report our RV data for $\mu$ UMa in Table~\ref{tab2}.

To check the long-term stability of the BOES, we have monitored the RV standard star $\tau$ Ceti since 2003.  RV measurements for this star show an rms scatter of $\sim$7 m s$^{-1}$ \citep{lee13}.

%__________________________________________________________________
%
\section{RADIAL VELOCITY VARIATIONS AND ORIGIN}
\subsection{Orbital solutions}
\begin{figure*}
   \centering
   \includegraphics[width=10cm]{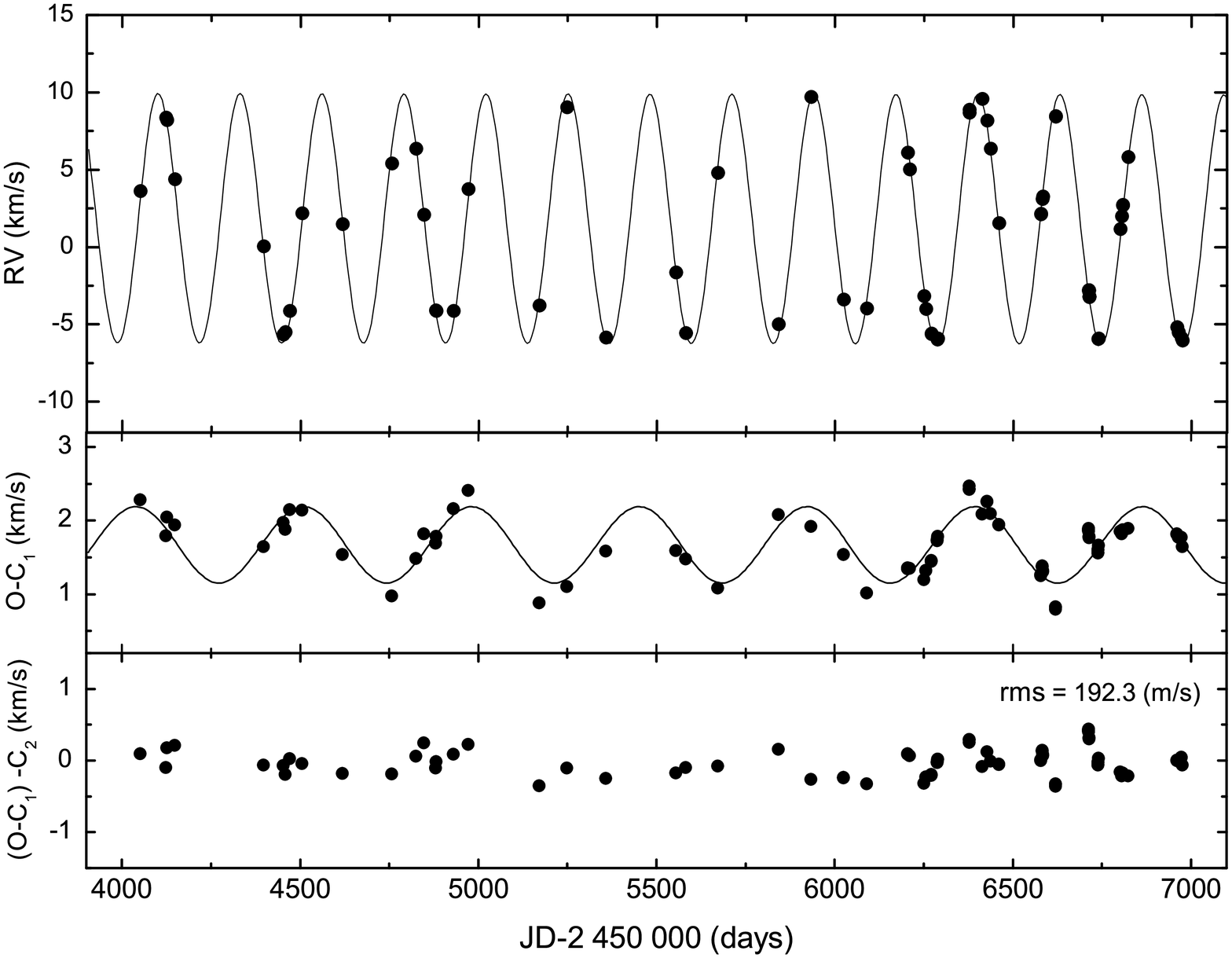}
      \caption{RV measurements for $\mu$ UMa from November 2006 to November 2014. (\emph{top panel}). Observed RVs for $\mu$ UMa and the orbital fit due to companion B. (\emph{middle panel}).
Orbital fit to the secondary variations. (\emph{bottom panel}) The residual velocities
after subtracting the two periodic signals.
              }
         \label{rv}
   \end{figure*}

In order to search for a periodicity in the RV data, we performed the Lomb-Scargle (hereafter, L-S) periodogram analysis \citep{sca82}.
The L-S periodogram of the RV measurements for $\mu$~UMa (\emph{top panel} in Fig.~\ref{rv}) shows a significant peak at a frequency of
0.00435~c~d$^{-1}$, corresponding to previously known spectroscopic binary period of 230.0 days. We have fitted (the solid lines in Fig.~\ref{rv}) and removed these variations.  The L-S analysis of the residuals shows a significant peak at a period of 471.2 days (\emph{middle panel} in Fig.~\ref{rv}). The L-S power of this peak corresponds to a false alarm probability (hereafter, FAP) of $<$  10$^{-6}$, adopting the procedure
described in \citet{cum04}.

\begin{figure}
   \centering
   \includegraphics[width=8cm]{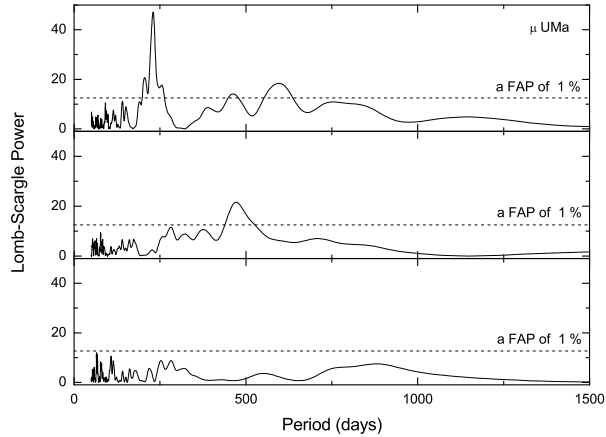}
      \caption{Periodogram of the RV measurements for $\mu$ UMa. (\emph{top panel}) The L-S periodogram of original data shows a significant power at a period of 230 days. (\emph{middle panel}) The same periodogram for the residuals after subtracting the main period. The largest peak is at a period of 471 days. (\emph{bottom panel}) Periodogram of the RV residuals after removing of two periods. The horizontal dotted lines indicate a FAP threshold of 1 $\times 10^{-2}$ (1\%).
      }
         \label{power}
   \end{figure}

Figure~\ref{power} (\emph{bottom panel}) shows the power spectrum of the velocity residuals after removing the two periodic signals. No significant variations are found: all residual peaks have a FAP threshold over 1$\times 10^{-2}$. An orbital fit yields an orbiting primary with a period $P$ = 230.0$\pm$ 0.1 days, a semi-amplitude $K$ = 7.88 $\pm$ 0.06 km s$^{-1}$, and an eccentricity $e$ = 0.02 $\pm$ 0.01.
The secondary  variations  have a period of 471.2 $\pm$ 2.1 days and a semi-amplitude of 0.52 $\pm$ 0.03 km~s$^{-1}$.

The mass of the $\mu$~UMa~A~ is $M_{A}$ = 2.2 $M_{\odot}$ \citep{mas08} yielding a companion mass $M_{B}$~sin~$i$ = 0.38 $M_{\odot}$. The semi-major axis of the system is 0.95 $\pm$ 0.01 AU. The low orbital inclination and its large error (13.6 $\pm$ 12.8\,$^{\circ}$) result in a large uncertainty in the mass of component B: $\mu$~UMa~B $=$ 1.6~$^{+25.7}_{-0.8}$ $M_{\odot}$.

\begin{figure}
   \centering
   \includegraphics[width=8cm]{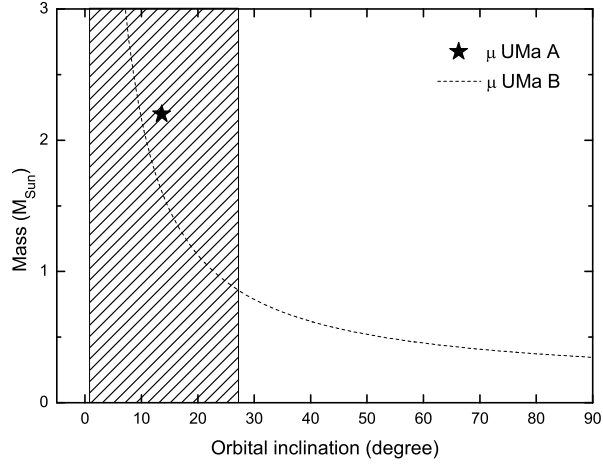}
      \caption{Companion mass for $\mu$ UMa~A with regard to the orbital inclination of the system. The
hatched  area indicates the variation range of inclination calculated by \citet{ren13} and the asterisk marks the position of $\mu$~UMa~A.
        }

        \label{inc}
   \end{figure}

Figure~\ref{inc} shows the companion mass for $\mu$~UMa A as a function of the orbital inclination of the system. The hatched region indicates the variation range in the orbital inclination and, thus, in the companion mass. $\mu$~UMa B has a nominal mass of $\approx$1.6 $M_{\odot}$  which is close to that of $\mu$~UMa A (2.2 $M_{\odot}$).
If $\mu$~UMa B were an evolved star with  high luminosity, then we should see its spectral features in our data.
However, no such traces were found in our spectra. Thus, $\mu$~UMa\,B is probably a less luminous dwarf star. Having a few times the solar luminosity, it would not give a noticeable effect on the spectrum or brightness of $\mu$~UMa (1148 $L_{\odot}$).

After removing the RV signals of 230 and 471 days, the dispersion of the RV residuals is 192.3 m s$^{-1}$, which is significantly higher than the
RV precision for the RV standard star $\tau$~Ceti ($\sim$7~m~s$^{-1}$) or the typical internal error of individual RV accuracy of $\sim$8~m~s$^{-1}$ of $\mu$~UMa. A periodogram of the RV residuals, however, does not show any additional periodic signal as shown in Figure~\ref{power} (\emph{bottom panel}).
This excess variability may arise from solar-like stellar oscillations. We estimate the amplitude of these using the scaling relationships of \citet{kje96}. The mass and luminosity of $\mu$~UMa A results in a velocity amplitude for the oscillations, $v_{osc}$ $\approx$120 m\,s$^{-1}$, or comparable to the residual scatter of our RV measurements. Since such variations would be stochastic, it is no surprise that we see no significant peaks in the periodogram of the final RV residuals.

Orbital elements for $\mu$ UMa B are listed in Table~\ref{tab3}. In the next subsections, we will test different assumptions about the nature of the second 471-day periodicity and the reasons for large scatter in the residuals from the two-period solution.

\subsection{The HIPPARCOS and COBE/DIRBE infrared photometries}
\begin{figure}
%\epsscale{.80}
\centering
\includegraphics[width=8cm]{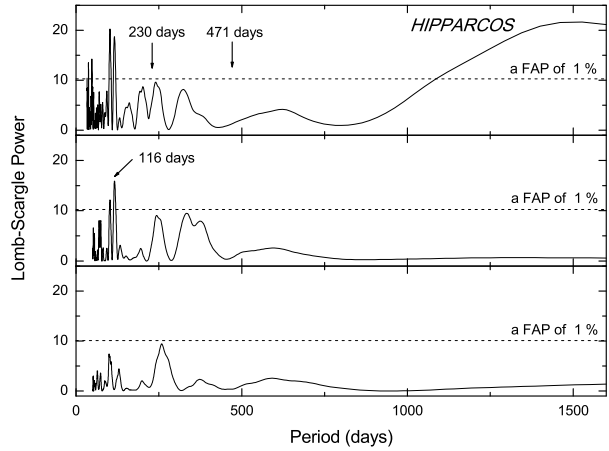}
\caption{The \emph{HIPPARCOS} photometric measurements. (\emph{top panel}) The L-S periodogram shows three significant peaks  at a periods of 102, 116, and 1506 days. The arrows mark the location of the orbital period of 230 and 471 days. (\emph{middle panel}) The same periodogram for the residual after subtracting the main period at 1506 days. The largest peak is at a period of 116 days. (\emph{bottom panel}) Periodogram of the RV residual after removing the two periods. The horizontal dotted lines indicate a FAP threshold of 1 $\times 10^{-2}$ (1\%).
   }
\label{hip}
\end{figure}

We analyzed the \emph{HIPPARCOS} photometry \citep{esa97} for $\mu$~UMa. For three years, between JD\,2447879 and JD\,2448978, the \emph{HIPPARCOS} satellite obtained 92 photometric measurements for $\mu$~UMa with an rms scatter of 0.0168 mag, corresponding to 0.53\% variations. Figure~\ref{hip} shows the L-S periodogram of these measurements. There are a few significant peaks near the periods of 102, 116, and 1506 days with FAPs of $<$ 5 $\times 10^{-4}$. After removing nominally the most significant period of 1506 days, we found a period of 116 days, irrelevant to the RV signals. The \emph{HIPPARCOS} photometry is unable to provide information on the possible origin of the RV variations.

\begin{figure}
\centering
\includegraphics[width=8cm]{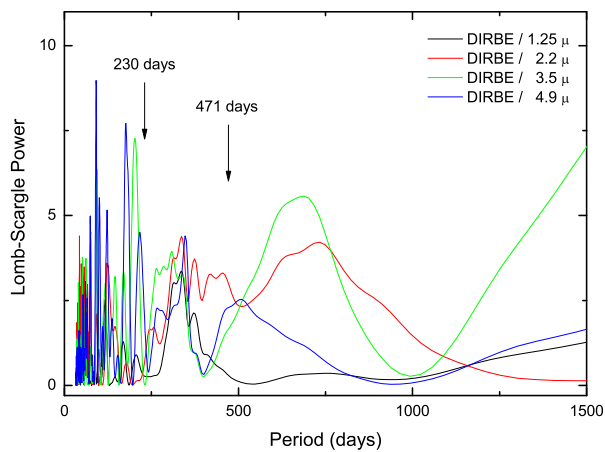}
\caption{The L-S periodograms of the \emph{COBE/DIRBE} satellite 1.25, 2.2, 3.5, and 4.9$\mu$ flux intensity measurements for $\mu$ UMa. The arrows mark the location of the periods at  230 and 471 days.
      }
\label{dirbe}
\end{figure}

Almost simultaneously, between JD\,2447875 and JD\,2449149, $\mu$ UMa was observed in the near-infrared (NIR) 1.25, 2.2, 3.5, and 4.9$\mu$ bands by  NASA's \emph{COBE} (Cosmic Background Explorer) satellite with the \emph{DIRBE} (Diffuse Infrared Background Experiment) instrument. The total number of 76 weekly averaged fluxes in each band were extracted for $\mu$~UMa from the \emph{COBE/DIRBE} archives \citep{pri10}. The L-S periodogram analysis of the 1.25, 2.2, 3.5, and  4.9$\mu$ fluxes does not reveal any significant signals in the domain of interest (Fig.~\ref{dirbe}).

\subsection{Line bisector variations}
Periodic RV variations can be produced by the rotational modulation of surface features or by pulsations. Surface features such as spots or non-radial pulsations cause an asymmetric, periodic distortion in the line profile, which can be detected by line bisector analysis. The difference in the bisectors of line widths between the top and bottom of the line profile is defined as the bisector velocity span (BVS). The changes in the spectral line bisector can also be quantified using the bisector velocity curvature (BVC).

\begin{figure}
%\epsscale{.80}
\centering
\includegraphics[width=8cm]{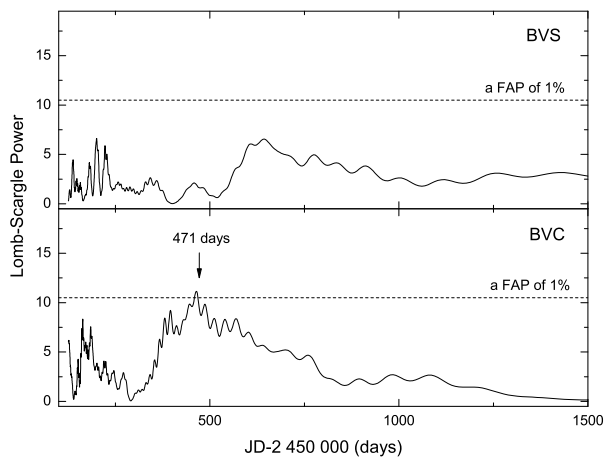}
\caption{The L-S periodograms of line bisector variations of the Ni I 6643.6 {\AA} spectral line for $\mu$ UMa. (\emph{top panel}) BVS measurements show no significant peak. (\emph{bottom panel}) BVC measurements indicate a significant power at a period of 463.6 days. The arrow marks the location of the secondary orbital period of 471 days and the horizontal dotted lines indicate a FAP threshold of 1 $\times 10^{-2}$ (1\%).
   }
\label{bvs1}
\end{figure}

In choosing the points defining the bisector span, it is important to avoid the wings and cores of the profile where the errors of the bisector measurements are large and noisy. To search for variations in the spectral line shapes for $\mu$~UMa, line bisector variations were computed for a strong and unblended spectral line with a high flux level, namely Ni I 6643.6 {\AA} as described in \citet{hat05} and \citet{lee13,lee14}. The selected line shows a high flux level and is located beyond the I$_{2}$ absorption region so that contamination should not affect our bisector measurements.
We estimated the bisector variations of the profile between two different flux levels at 0.8 and 0.4 of the central depth as the span points. The L-S periodograms of the line bisector variations for $\mu$~UMa are shown in Fig.~\ref{bvs1}. Even though the BVS does not show any obvious peak, the BVC indicates a peak near the period of 463.6 days with a FAP of $\sim$$10^{-3}$. This suggests that RV variations are accompanied by line-shape changes.
%(BVS = V$_{T}$ $-$ V$_{B}$)
%(BVC = [V$_{T}$ $-$ V$_{C}$] $-$ [V$_{C}$ $-$ V$_{B}$])
%A BVS is simply the velocity difference in the bisectors of line widths between the top and bottom of the line profile (BVS) and a BVC is the difference of the velocity span of the upper half and the lower half of the bisector (BVC).

%
\subsection{Chromospheric activities}
Magnetic fields, which are generated by turbulence in the outer convection zone, cause a very broad range of surface phenomena, such as sunspots, plages, and flares. Stellar activity refers to similar features that occur in late-type stars that have an outer convection zone. Thus, the nature of stellar activity is related to the existence and depth of an outer convection zone. The depth depends on the spectral type: F type stars have shallow convection zones while middle M type stars are totally convective.
Substitution of basic stellar parameters for $\mu$ UMa A (see Table\,\ref{tab1}) into \citet{fre02}'s calibration of  $x_{gran}$/$R_{\star}$ $\approx$ 0.0025 $\times$ ($R_{\star}$/$R_{\odot}$)~($T_{\rm{eff},\star}$/$T_{\rm{eff},\odot}$)~($M_{\odot}$/$M_{\star}$) yields an  expected sizes for the surface granulation of $\approx$0.378 $R_{\star}$, approximately corresponding to eight large convection cells across the visible equator of the star.

Frequently used optical chromospheric activity indicators are the EW variations of Ca II H \& K, H$_{\beta}$, Mg~I~b triplet, He~I~D$_{3}$,
Na~I~D$_{1}$ \& D$_{2}$, H$_{\alpha}$, and Ca II infrared triplet (IRT) lines. The behaviours of the different optical chromospheric activity indicators reflect the atmospheric condition at different atmospheric heights: Na I D$_{1}$ \& D$_{2}$ and Mg~I~b triplet for the upper photosphere and lower chromosphere, Ca~II~IRT lines for the lower chromosphere, H$_{\alpha}$, H$_{\beta}$, Ca~II~H \& K for the middle chromosphere and He I D$_{3}$ for the upper chromosphere.

Of these, the most common indicator of chromospheric activity is the well-known S-index of the Ca II H \& K lines \citep{vau78}, which is well studied for F to K type stars.
The Ca~II~H \& K line region for $\mu$ UMa obtained by the BOES does not have a sufficient S/N ratio to estimate EW variations.
Another Ca~II~indicator, the Ca~II~IRT lines, has also been measured \citep{lar93,hat03,lee13}. Changes in the core profile of the Ca II IRT reflect variations in stellar chromospheric activity and qualitatively related to the variations in the Ca II H \& K flux \citep{lar93}. However, it also is not suitable because saturation of our CCD spectra is seen at wavelengths longer than around 6800~{\AA}. The Na I D$_{1}$ \& D$_{2}$ and Mg~I~b triplet lines are excluded from the study because they are located inside I$_{2}$ molecular region.

\begin{figure}
   \centering
   \includegraphics[width=8cm]{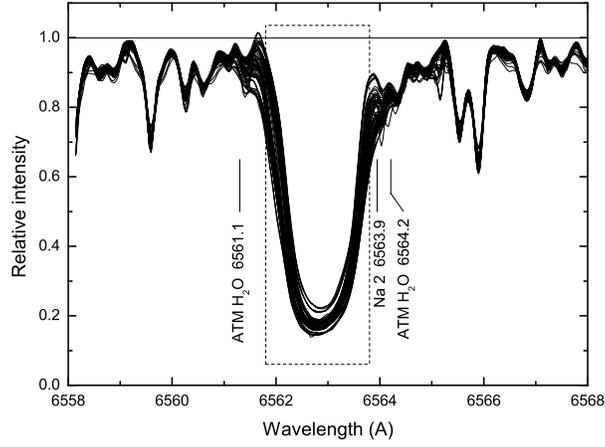}
      \caption{Line profile near the H$_{\alpha}$ region for $\mu$ UMa. A square box denotes the range of the H$_{\alpha}$ in which the EWs were measured.
        }
        \label{h_a}
   \end{figure}

\begin{figure}
   \centering
   \includegraphics[width=8cm]{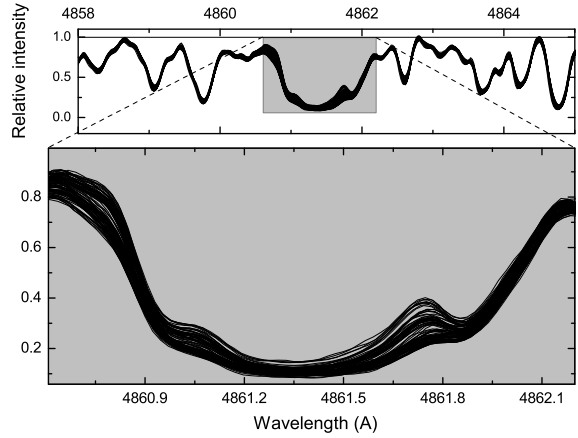}
      \caption{Line profile near the H$_{\beta}$ region for $\mu$ UMa. A square gray box denotes the range of the H$_{\beta}$ in which the EWs were measured.
        }
        \label{h_b}
   \end{figure}

\begin{figure}
   \centering
   \includegraphics[width=8cm]{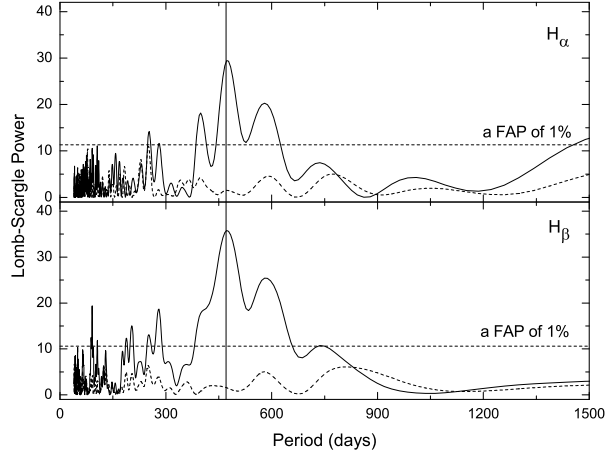}
      \caption{The L-S periodograms of the H$_{\alpha}$ and H$_{\beta}$ EW variations for $\mu$~UMa. The solid lines are the Lomb-Scargle periodogram of the H$_{\alpha}$ EW (\emph{top panel}) and the H$_{\beta}$ EW (\emph{bottom panel}), which indicate significant peaks at a period of $\sim$473 days. The dashed lines show the periodogram of the residual after removing of the main period from the original data. The vertical dashed line marks the location of the period of 471 days and the horizontal dotted lines indicate a FAP threshold of 1 $\times 10^{-2}$ (1\%).
        }
        \label{h_power}
   \end{figure}

In this work, we thus selected the H$_{\alpha}$ and H$_{\beta}$ lines as chromospheric indicators. The H$_{\alpha}$ is believed to have a strong correlation with the Ca II index \citep{gia89,rob90,str90} and is sensitive to the atmospheric stellar activity \citep{kur03}. It is often used as a chromospheric indicator \citep{lab86,pas91,tha93,mon95}. The H$_{\beta}$ absorption line was also measured, which originates in the middle layers of stellar atmosphere and is sensitive to the stellar activity.
In Figure~\ref{h_a} and \ref{h_b}, the shapes of the H$_{\alpha}$ and H$_{\beta}$ EW variations are superimposed on each other.
We measured the EWs using a band pass of $\pm$ 1.0 ${\AA}$ for H$_{\alpha}$ and $\pm$ 0.8 ${\AA}$ for H$_{\beta}$ centered on the core of the lines to avoid nearby blending lines and ATM H$_{2}$O absorption lines. The L-S periodograms of the H$_{\alpha}$ and H$_{\beta}$ EW variations for $\mu$~UMa are shown in Fig.~\ref{h_power}. Both periodograms show a large power at $\sim$473 days, close to the RV period of 471.2 days.

\section{DISCUSSION}
We found long-term RV variations in the spectroscopic binary M giant $\mu$~UMa. Two periodic signals are present in the data, one at 471.2 days and another at 230.0 days. The 230-d period is due to a known stellar companion. The nature of the longer period variations was assessed using ancillary data.

Line bisector variations may arise from variations of the instrumental profile (IP)  as well as stellar variations. Generally, the line bisector variations of IP can be significant, but they are typically much lower than the errors in the line bisectors determined from the stellar spectra. While our bisector analysis showed that the amplitude variations are smaller than those of the RV, it revealed variations of $\sim$45.8 m s$^{-1}$ (BVS) and of $\sim$65.0 m s$^{-1}$ (BVC).
The BOES exhibits a slightly unstable IP variation due to the possible mechanical, thermal and ambient air pressure instabilities from season to season. We examined the stability of the instrumental precision by measuring the  bisectors on the reference star $\tau$ Ceti. Bisector measurements taken on the same, or consecutive nights (14 spectra during JD 2456552.2059 and 24556555.3222) show BVS variations of $\sim$39 m s$^{-1}$ and BVC variations of $\sim$35~m~s$^{-1}$. Considering that the long-term precision is probably worse, the bisector variations due to instrumental effects are comparable to the variations measured for $\mu$~UMa. Normally this would indicate no significant bisector variations in $\mu$~UMa; however, BVC variations are seen with a period of 463 days, very close to the secondary period of 471 days seen in RV variations.
We note that the instrumental BVC variations are about a factor of two smaller than the BVC variations in $\mu$~UMa, so the BVC variations seen in $\mu$~UMa  are probably real.

Stellar activity, such as spots, plage, and filaments, can induce RV variations that can mask or even mimic the RV signature of orbiting companions. Since we were unable to use the Ca II H \& K lines as indicators due to the poor S/N ratio of our spectra in this
wavelength region, we selected hydrogen lines for the activity indicators.
We found that $\mu$~UMa shows a prominent power in the EW variations of the H$_{\alpha}$ and H$_{\beta}$ lines, possibly indicating a high level of stellar activity. Two hydrogen lines show  period ($\sim$473 days) consistent to the 471-d period seen in the RVs of $\mu$~UMa.
Hydrogen lines can be signatures of stellar chromospheric activity and H$_{\alpha}$ surveys have shown that the velocity structure of the chromosphere is reflected in the H$_{\alpha}$ profile. In particular, the H$_{\alpha}$ line is sensitive to chromospheric structure and conditions in  cool giant and supergiant stars \citep{cra85,smi89}.
Therefore, the secondary RV period of 471 days may very well be due to chromospheric activity.

The stellar rotational period can be important in identifying the origins of RV variations \citep{lee08,leea12}.
From the measured projected rotational velocities, the stellar radius, and the stellar inclination, we can derive  the range for the stellar rotational period. Assuming that the stellar inclination is the same as the orbital inclination (i.e. aligned angular momentum vectors),
the stellar rotational period is

   $P_{rot} = 2 \pi$$R_{\star}$/[$v_{rot}$ sin($i$)] = 118~$^{+107}_{-111}$ days.\\

\hskip -15pt
Interestingly, the \emph{HIPPARCOS} photometry shows a significant period at 116 days, consistent
with the estimated rotational period.

There seems to be a discrepancy between the estimated stellar rotational period and the 471-d RV period. The latter is found in activity indicators  which nominally indicates rotational modulation at the stellar rotational period. There are several ways to resolve this discrepancy.

First, the stellar spin axis of the star may not be aligned with the orbital axis. Rossiter-McLaughlin measurements of eclipsing binaries have found  such mis-aligned systems \citep{alb14}. If the stellar spin axis actually has an inclination of $\approx$70$^\circ$, then the rotational and RV period are consistent. Second, the 116-d period found in the \emph{HIPPARCOS} photometry may actually be a harmonic of the
true rotational period. One-fourth of the 471-d RV period is 118 d. Possibly the spot distribution at the time of the \emph{HIPPARCOS} measurements consisted of four spots equally spaced in longitude. This would result in one-fourth of the true rotational period. Finally, it may be that the
471-d period is actually not the rotation period of the star.

Using $L_{\odot}$ = 1148 and $T_{\rm{eff}}$ = 3899~K for $\mu$~UMa A \citep{mas08}, we can locate it in the Hertzsprung-Russell (H-R) diagram at the domain of an Asymptotic Giant Branch (AGB) star.
Substitution of masses of $\mu$~UMa A and $\mu$~UMa B to the Eggleton's approximate formulae \citep{egg83} yields the Roche lobe radius $R_L$ = 0.40 in units of orbital separation or equal to 0.39 AU. The radius of $\mu$~UMa A  is  $R_{*}$ = 74.7$R_{\odot}$ or 0.347 AU. This turns out that $\mu$~UMa A is an AGB star which is close to filling its Roche lobe.

Pulsations in AGB stars are quit common phenomenon and there is still a mysterious ``Sequence D'' group among them, showing a long secondary period (LSP; $>$~400 days) that follow a period-luminosity relation \citep{woo99}. A ``Sequence D'' star show  evidence of warm chromosphere  through strong and variable H$_{\alpha}$ line absorption that varies in phase with LSP  \citep{woo04}.
We suggest that the nature of the long secondary period of $\approx$471 days and variations in the hydrogen lines may relate $\mu$~UMa A closely to the ``Sequence D'' variable AGB stars, although it does not show short-period pulsations. The origin of existence of ``Sequence D'' AGB stars is yet to be explained. According to \citet{woo04}, the most likely explanation is that the LSPs result from a low degree $g^{+}$ modes confined to the outer layer of the red giant, combined with a large-scale stellar spot activity that gives rise to an observed chromosphere and the irregularity of the light curve. \citet{sai15} found from the theoretical models that ``Sequence D'' period-luminosity relation is roughly consistent with the predictions for dipole oscillatory convective modes in AGB models.

\begin{figure}
   \centering
   \includegraphics[width=8cm]{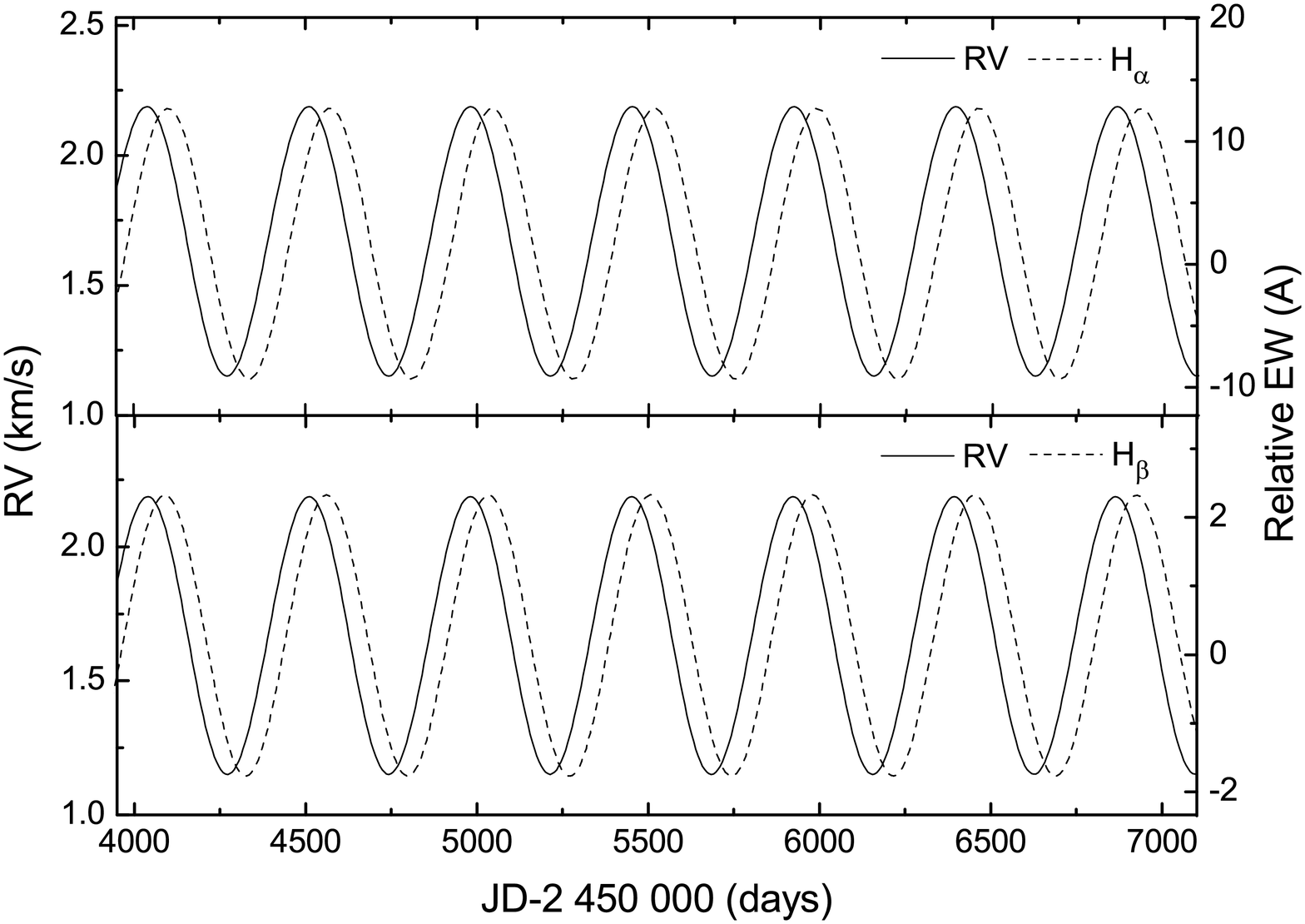}
      \caption{Schematic of the secondary RV and the hydrogen line EW variations vs. time. RV measurements vs. the H$_{\alpha}$ EW (\emph{top panel}) and the H$_{\beta}$ EW variations (\emph{bottom panel}). The solid lines are RV curve and the dashed lines show the EW measurements of the hydrogen lines. The
scale of the RV and the EW were adjusted arbitrarily to match the two curves.
        }

        \label{h_ew+rv1}
   \end{figure}

Figure~\ref{h_ew+rv1} shows schematics of the secondary RV of $\mu$~UMa  and the H$_{\alpha}$ EW and H$_{\beta}$ EW variation cycle of the host star as a function of the time. The EW variations show a time delay of $\sim$90 days (0.19 phase) relative to the RV variations.
For completeness, we would like to mention that  pulsations of extended atmospheres of an AGB star can also explain the 0.19 phase lag
between the RV and the EW variations of hydrogen lines and the metallic line bisector variations in $\mu$~UMa.  RVs are the mirror image of the optical flux variations in the pulsating stars. Thus, in the case of pulsations in $\mu$~UMa, the EW maximum (see Figure\,10) should precede the optical flux variations by about 0.25 phase. Earlier,  \citet{leeb12} detected a phase lag between the 706-day secondary variations in the H$_{\alpha}$ EW and the 1.25$\mu$ flux in the exoplanet host K giant HD~66141. The H$_{\alpha}$ line EW variations in HD~66141 precede the 1.25$\mu$ flux maximum by 0.18 phase, very close to the observed value found
in $\mu$~UMa. Observations of other M giants and Mira variables at different wavelengths show phase lags  with the EW  maximum appearing about 0.18 phase before the maximum at 1.25$\mu$ \citep{pri10}. Such phase lags in Oxygen-rich Miras are likely due to titanium oxide (TiO) variability \citep{alv98}.

Considering all these lead us to conclude that the RV and the EW variations in the spectroscopic binary M giant $\mu$~UMa A originate from complex pulsations and the chromospheric activity.
Additional  RV measurements with contemporaneous photometric measurements will confirm the nature of secondary long-period variations for this star.

\acknowledgments
BCL acknowledges partial support by the KASI (Korea Astronomy and Space Science Institute) grant 2015-1-850-04. Support for MGP was provided by the National Research Foundation of Korea to the Center for Galaxy Evolution Research (No. 2012-0027910). DEM acknowledges his work as part of the research activity of the National Astronomical Research Institute of Thailand (NARIT), which is supported by the Ministry of Science and Technology of Thailand. APH acknowledges grant HA 3279/8-1 from the Deutsch Forschungsgemeinschaft (DFG). This research made use of the SIMBAD database, operated at the CDS, Strasbourg, France. We thank the developers of the Bohyunsan Observatory Echelle Spectrograph (BOES) and all staff of the Bohyunsan Optical Astronomy Observatory (BOAO).

\clearpage

%-------------------------------------------------------------
%
%-------------------------------------------------------------
%
%-----------------------------------------------------------
%
%-----------------------------------------------------------
%
%-----------------------------------------------------------
%
%-----------------------------------------------------------
%
%-----------------------------------------------------------
%
%-----------------------------------------------------------
%
%-----------------------------------------------------------
%
%-----------------------------------------------------------
\clearpage

%-------------------------------------------------------------

\begin{deluxetable}{lcc}
\tablewidth{0pt}
\tablecaption{Stellar parameters for $\mu$ UMa analyzed in the present paper.
\label{tab1}}
\tablehead{
\colhead{Parameter}                           & \colhead{Value}      & \colhead{Reference}  }
\startdata
    Spectral type\dotfill                    & M0 III SB             & 1  \\
    $\textit{$m_{v}$}$ (mag)\dotfill         & 3.1544 $\pm$ 0.0026   & 1  \\
    $\textit{B-V}$ (mag)\dotfill             & 1.603 $\pm$ 0.006     & 2  \\
%    age                &[Gyr]& 1.85 $\pm$ 0.34 &  Derived\tablefootmark{a}  \\
    $\emph{d}$  (pc)\dotfill                 & 76 $\pm$ 4            & 3  \\
    RV (km s$^{-1}$)\dotfill                 & $-$ 20.4              & 2  \\
    Parallax (mas)\dotfill                   & 14.16 $\pm$ 0.54      & 3  \\
    $T_{\rm{eff}}$ (K)\dotfill               & 3899 $\pm$ 35         & 3  \\
    $\rm{[Fe/H]}$ (dex)\dotfill              & $-$ 0.04              & 4  \\
    log $\it g$ (cgs)\dotfill                & 1.0                   & 3  \\
    $v_{\rm{micro}}$ (km s$^{-1}$)\dotfill        & 2.01 $\pm$ 0.3   & 2  \\
                                                  & 1.5  $\pm$ 0.1   & 4  \\
    $\textit{$R_{\star}$}$ ($R_{\odot}$)\dotfill  & 74.7             & 3  \\
%                            &             & xx $\pm$ x.x  &  Derived\tablefootmark{a} \\
    $\textit{$M_{\star}$}$ ($M_{\odot}$)\dotfill  & 2.2              & 3  \\
    $\textit{$L_{\star}$}$ ($L_{\odot}$)\dotfill  & 1148             & 3  \\
    $v_{\rm{rot}}$ sin $i$ (km s$^{-1}$)\dotfill  & 7.5              & 3  \\
    Inclination $\emph{i}$ ($^{\circ}$)\dotfill   & 13.6 $\pm$ 12.8  & 5  \\
\enddata
%\tablenotetext{a}{Star LP 608--62 is also known as BD+1\arcdeg 2341p.  We will make this footnote extra long so that it extends over two lines.}
\tablerefs{
(1) \citet{esa97}; (2) \citet{van07}; (3) \citet{mas08}; (4) \citet{and12}; (5) \citet{ren13}}
\end{deluxetable}

\clearpage
\begin{landscape}
\begin{deluxetable}{crc|crc|crc|crc}
\tabletypesize{\scriptsize}
%\tablewidth{11pc}
\tablewidth{0pt}

\tablecaption{RV measurements for $\mu$ UMa from November 2006 to November 2014.
\label{tab2}}
\tablehead{\colhead{JD} & \colhead{RV} & \colhead{$\pm \sigma$} &
            \colhead{JD} & \colhead{RV} & \colhead{$\pm \sigma$} &
            \colhead{JD} & \colhead{RV} & \colhead{$\pm \sigma$} &
            \colhead{JD} & \colhead{RV} & \colhead{$\pm \sigma$}  \\
           \colhead{$-$2,450,000} & \colhead{km\,s$^{-1}$} & \colhead{m\,s$^{-1}$} &
           \colhead{$-$2,450,000} & \colhead{km\,s$^{-1}$} & \colhead{m\,s$^{-1}$} &
           \colhead{$-$2,450,000} & \colhead{km\,s$^{-1}$} & \colhead{m\,s$^{-1}$} &
           \colhead{$-$2,450,000} & \colhead{km\,s$^{-1}$} & \colhead{m\,s$^{-1}$}}
\startdata
4051.3733 &    3.6352  &  7.6  &  6204.3504 &   6.1129  &  8.0  &  6437.0216  &   6.3665  &  6.6 &  6738.0842  &  $-$5.9354  & 9.7  \\
4123.1346 &    8.3559  &  5.4  &  6204.3522 &   6.1098  &  7.9  &  6459.9793  &   1.5398  &  6.6 &  6739.1190  &  $-$5.9416  & 9.2  \\
4126.1527 &    8.2126  &  5.0  &  6204.3538 &   6.1184  &  8.9  &  6459.9819  &   1.5460  &  7.0 &  6739.1205  &  $-$5.9462  & 9.3  \\
4147.1043 &    4.3964  &  5.7  &  6210.3164 &   5.0220  &  8.1  &  6578.3190  &   2.1344  &  8.4 &  6739.1215  &  $-$5.9479  & 8.6  \\
4396.3722 &    0.0593  &  8.2  &  6210.3183 &   5.0206  &  8.6  &  6578.3207  &   2.1243  &  9.2 &  6739.9376  &  $-$5.9168  & 9.9  \\
4452.4213 & $-$5.6589  &  8.5  &  6250.3813 & $-$3.1611  &  7.6  &  6578.3219  &   2.1375  &  10.0 &  6739.9390  &  $-$5.9121  & 8.6  \\
4458.4272 & $-$5.4896  &  7.8  &  6250.3836 & $-$3.1597  &  8.9  &  6578.3231  &   2.1383  &  9.2 &  6739.9402  &  $-$5.9125  & 9.4  \\
4470.4113 & $-$4.1356  &  6.9  &  6250.3855 & $-$3.1600  &  8.4  &  6582.3058  &   3.1038  &  9.0 &  6801.0235  &   1.1607  & 6.4  \\
4505.1241 &    2.1730  &  7.4  &  6256.3767 & $-$3.9975  &  7.2  &  6582.3072  &   3.1168  &  8.0 &  6805.0243  &   2.0000  & 7.0  \\
4618.9881 &    1.4773  &  7.8  &  6256.3779 & $-$3.9927  &  7.3  &  6582.3084  &   3.1087  &  9.2 &  6805.0263  &   1.9923  & 7.3  \\
4756.3428 &    5.4204  &  11.7 &  6256.3795 & $-$3.9922  &  7.4  &  6583.3497  &   3.2820  &  9.1 &  6805.0277  &   1.9975  & 7.5  \\
4824.1808 &    6.3548  &  7.2  &  6271.3977 & $-$5.6086  &  7.8  &  6583.3514  &   3.2680  &  9.0 &  6808.0766  &   2.7068  & 9.0  \\
4847.3822 &    2.0807  &  7.4  &  6271.4004 & $-$5.6130  &  7.6  &  6583.3526  &   3.2694  &  8.2 &  6808.0780  &   2.6977  & 7.1  \\
4880.2588 & $-$4.1102  &  7.6  &  6271.4030 & $-$5.6214  &  7.4  &  6583.3538  &   3.2617  &  9.8 &  6808.0792  &   2.7039  & 8.0  \\
4881.0774 & $-$4.1269  &  5.9  &  6287.2706 & $-$5.9625  &  8.2  &  6620.2082  &   8.4280  &  6.8 &  6808.0804  &   2.7124  & 8.5  \\
4930.1547 & $-$4.1319  &  7.1  &  6287.2723 & $-$5.9713  &  7.9  &  6620.2143  &   8.4529  &  9.9 &  6822.9835  &   5.8350  & 7.2  \\
4971.0400 &    3.7547  &  4.9  &  6287.2742 & $-$5.9661  &  8.4  &  6620.2207  &   8.4609  &  10.8 &  6822.9857  &   5.8345  & 6.1  \\
5171.2614 & $-$3.7913  &  7.7  &  6288.3702 & $-$5.9115  &  7.6  &  6712.1585  &  $-$2.8174  &  12.1 &  6822.9876  &   5.8342  & 6.7  \\
5248.2100 &    9.0511  &  8.6  &  6288.3721 & $-$5.9133  &  7.8  &  6712.1596  &  $-$2.7897  &  13.0 &  6960.3097  &  $-$5.1749  & 8.5  \\
5356.9884 & $-$5.8384  &  7.4  &  6288.3740 & $-$5.9081  &  8.3  &  6712.1604  &  $-$2.8203  &  10.2 &  6960.3123  &  $-$5.1743  & 8.5  \\
5554.3643 & $-$1.6415  &  8.0  &  6377.0117 &   8.6963  &  6.6  &  6712.1613  &  $-$2.7977  &  11.1 &  6960.3149  &  $-$5.1778  & 8.5  \\
5581.1677 & $-$5.5462  &  9.0  &  6377.0139 &   8.6951  &  6.1  &  6712.1621  &  $-$2.8039  &  12.7 &  6964.3288  &  $-$5.5119  & 8.2  \\
5672.1456 &    4.7859  &  9.9  &  6378.0382 &   8.8768  &  6.4  &  6712.1638  &  $-$2.7874  &  10.6 &  6964.3365  &  $-$5.5088  & 7.9  \\
5842.3452 & $-$4.9747  &  7.3  &  6378.0411 &   8.8769  &  5.8  &  6714.0990  &  $-$3.2084  &  11.0 &  6964.3440  &  $-$5.5215  & 7.2  \\
5933.3511 &    9.7137  &  8.0  &  6413.0068 &   9.5935  &  7.4  &  6714.1013  &  $-$3.2204  &  8.7 &  6972.2647  &  $-$5.8781  & 9.0  \\
6024.0145 & $-$3.3994  &  8.0  &  6413.0101 &   9.5959  &  7.6  &  6714.1032  &  $-$3.2039  &  9.5 &  6972.2666  &  $-$5.8779  & 9.1  \\
6024.0172 & $-$3.4042  &  7.7  &  6427.1230 &   8.1729  &  6.1  &  6738.0805  &  $-$5.9402  &  8.3 &  6972.2684  &  $-$5.8745  & 9.1  \\
6088.9801 & $-$3.9817  &  8.7  &  6427.1248 &   8.1785  &  5.6  &  6738.0825  &  $-$5.9418  &  9.1 &  6975.2163  &  $-$6.0470  & 9.3  \\
\enddata
%\tablenotetext{a}{Includes scale factors described in the text.}
\end{deluxetable}
\end{landscape}

\clearpage

%-------------------------------------------------------------

\begin{deluxetable}{lc}
\tablewidth{0pt}
\tablecaption{Orbital parameters for $\mu$ UMa B.
\label{tab3}}
\tablehead{
\colhead{Parameter}                   & \colhead{$\mu$ UMa B}  }
\startdata
    Orbital period P (days)\dotfill                   & 230.0  $\pm$ 0.1       \\
    Velocity amplitude $\it{K}$ (km s$^{-1}$)\dotfill & 7.88  $\pm$ 0.06       \\
    $\it{e}$ \dotfill                                 & 0.02 $\pm$ 0.01        \\
    $\omega$ (deg)\dotfill                            & 338.5 $\pm$ 7.6        \\
    $\it T$$_{\rm{periastron}}$  (JD)\dotfill         & 2454087.5 $\pm$ 6.6    \\
    $m$  ($M_{\odot}$)\dotfill                        & 1.6~$^{+25.7}_{-0.8}$  \\
    $\it{a}$ (AU)\dotfill                             & 0.95  $\pm$ 0.01       \\
%    slope (km s$^{-1}$ year$^{-1}$)\dotfill             & 3.4 $\times$ 10$^{-2}$  \\
%    slope (km s$^{-1}$ day$^{-1}$)\dotfill             & 9.3 $\times$ 10$^{- 2}$  \\
    rms (km s$^{-1}$)\dotfill                          & 0.192                 \\
\enddata
%\tablenotetext{a}{Star LP 608--62 is also known as BD+1\arcdeg 2341p.  We will make this footnote extra long so that it extends over two lines.}
%\tablerefs{
%(1) Barbuy, Spite, \& Spite 1985; (2) Bond 1980; (3) Carbon et al. 1987}
\end{deluxetable}

\end{document}